\documentclass{emulateapj}
\usepackage{graphicx}
\usepackage{amssymb,amsfonts,amsmath,amstext,amsgen,amsopn,amsxtra,indentfirst,times}

\begin{document}

\title{A Non-isothermal Theory for Interpreting Sodium Lines in Transmission Spectra of Exoplanets}
\author{Kevin Heng\altaffilmark{1}, Aur\'{e}lien Wyttenbach\altaffilmark{2}, Baptiste Lavie\altaffilmark{1,2}, David K. Sing\altaffilmark{3}, David Ehrenreich\altaffilmark{2}, Christophe Lovis\altaffilmark{2}}
\altaffiltext{1}{University of Bern, Physics Institute, Center for Space and Habitability, Sidlerstrasse 5, CH-3012, Bern, Switzerland.  Email: kevin.heng@csh.unibe.ch}
\altaffiltext{2}{Observatoire de l'Universit\'{e} de Gen\`{e}ve, 51 chemin des Maillettes, 1290, Sauverny, Switzerland}
\altaffiltext{3}{Astrophysics Group, School of Physics, University of Exeter, Stocker Road, Exeter, EX4 4QL, UK}

\begin{abstract}
We present a theory for interpreting the sodium lines detected in transmission spectra of exoplanetary atmospheres.  Previous analyses employed the isothermal approximation and dealt only with the transit radius.  By recognising the absorption depth and the transit radius as being independent observables, we develop a theory for jointly interpreting both quantities, which allows us to infer the temperatures and number densities associated with the sodium lines.  We are able to treat a non-isothermal situation with a constant temperature gradient.  Our novel diagnostics take the form of simple-to-use algebraic formulae and require measurements of the transit radii (and their corresponding absorption depths) at line center and in the line wing for both sodium lines.  We apply our diagnostics to the HARPS data of HD 189733b, confirm the upper atmospheric heating reported by \cite{huitson12}, derive a temperature gradient of $0.4376 \pm 0.0154$ K km$^{-1}$ and find densities $\sim 1$--$10^4$ cm$^{-3}$.
\end{abstract}


\section{Introduction}

Due to its large cross section and favorable wavelength range, the sodium doublet lines have been a boon to astronomers seeking to characterise exoplanetary atmospheres \citep{ss00,brown01}.  In fact, the first detection of an exoplanetary atmosphere (that of HD 209458b) was accomplished via measuring the sodium doublet \citep{char02}.  Ever since that discovery, sodium has been detected in several hot Jupiters using both space- and ground-based transmission spectroscopy \citep{redfield08,sing08a,sing08b,snellen08,vm11,wood11,huitson12,sing12,zhou12,pont13,nikolov14,burton15,wy15}.

As better instruments come online and our ability to resolve the sodium lines improves, it is worth revisiting and redeveloping a theory of how to interpret them.  From a remote sensing perspective, a pair of fully resolved sodium lines can, in principle, map out the temperature-pressure profile of an atmosphere at high altitudes and yield the sodium abundance.  A quantity used to interpret absorption lines is the equivalent width \citep{spitzer78,draine11},
\begin{equation}
\begin{split}
W &= \int W_\lambda ~d\lambda, \\
W_\lambda &= 1 - \exp{\left( -\tau \right)},
\end{split}
\end{equation}
where $\tau$ is the optical depth of the intervening material and $\lambda$ denotes the wavelength.  The equivalent width is essentially the width of a box with the same depth as the trough of the absorption line, such that it encompasses the same area.  

While it is not immediately obvious, $W_\lambda$ and the transit radius ($R$) are actually independent observables.  A simple thought experiment demonstrates this.  Consider a fictitious star that emits achromatically.  At the exact moment of transit, one may record an absorption spectrum of the exoplanetary atmosphere.  If the wavelength coverage is sufficient, the sodium doublet and the continuum and thus $W_\lambda$ may be measured.  To record the transit radius requires temporal information: the change in flux in and out of transit.  In short, $W_\lambda$ measures changes in absorption across wavelength, while $R$ derives from the change in flux across time.  Certainly, stars are not achromatic light sources and one needs to measure the stellar lines in and out of transit to properly subtract out their influence, but this is an observational, rather than a theoretical, obstacle.

\begin{figure}[!t]
\begin{center}
\vspace{-0.1in}
\includegraphics[width=\columnwidth]{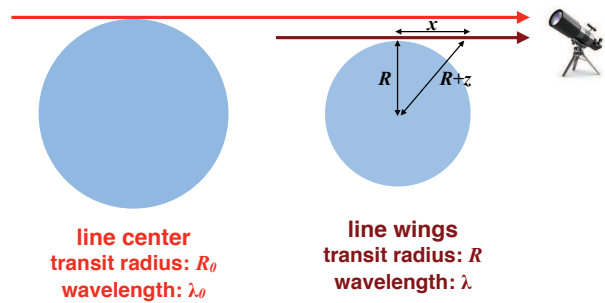}
\end{center}
\vspace{-0.2in}
\caption{Schematic of transit radii associated with the line center and wings of the sodium doublet.  As each line is expected to have the largest cross section at line center, the corresponding transit radius ($R_0$) is the largest.  In the line wings, where the cross section is the smallest, the transit radius ($R$) is correspondingly smaller.  The cartoon on the right reproduces the schematic depicted in \cite{fortney05}.}
\label{fig:schematic}
\end{figure}

Nevertheless, a theoretical challenge with interpreting $W$, for transmission spectra of exoplanets, is that the optical depth depends on the transit radius and the radius itself depends on wavelength (Figure \ref{fig:schematic}).  It is generally difficult to derive the functional form of $R(\lambda)$, but it is straightforward to measure it.  Additionally, while $W$ is associated with a fixed sightline in traditional studies of the interstellar medium, it is associated with a \textit{set of sightlines} in transmission spectra, each corresponding to a different wavelength.  Instead of a theory for $W$, we develop one for $W_\lambda(R)$.  Observationally, it takes the form,
\begin{equation}
W_\lambda = \frac{F_c - F}{F_c},
\end{equation}
where $F_c$ is the flux associated with the continuum and $F$ is the flux at any point within the line.  More specifically, both $F_c$ and $F$ are integrated over a (small) wavelength interval $\delta \lambda$ that is chosen based on practical constraints.  We shall term $F/F_c$ the \textit{absorption depth} (and leave $W_\lambda$ nameless).

\section{Theory}

\subsection{Order-of-Magnitude Estimates}

Since the line-center wavelength of both sodium lines is $\lambda_0 \approx 0.6$ $\mu$m, the line-center frequency is $\nu_0 = c/\lambda_0 \approx 5 \times 10^{14}$ Hz.  This means that the Doppler line width is
\begin{equation}
\begin{split}
\Gamma_{\rm D} &= \frac{\nu_0}{c} \left( \frac{2 k_{\rm B} T}{m} \right)^{1/2} \\
&\approx 5 \times 10^9 \mbox{ Hz} \left( \frac{T}{10^3 \mbox{ K}} \right)^{1/2} \left( \frac{m}{2 m_{\rm H}} \right)^{-1/2},
\end{split}
\end{equation}
where $c$ is the speed of light, $k_{\rm B}$ is the Boltzmann constant, $T$ is the temperature, $m$ is the mean molecular mass and $m_{\rm H}$ is the mass of the hydrogen atom.  We assume that sodium is a trace element in the atmosphere and that Doppler broadening is mediated by a dominant buffer or inert gas (i.e., molecular hydrogen).

By contrast, the natural line width is 
\begin{equation}
\Gamma_{\rm L} = \frac{A_{21}}{2\pi} \approx 10^7 \mbox{ Hz},
\end{equation}
where $A_{21} \approx 6 \times 10^7$ s$^{-1}$ is the Einstein A-coefficient.  Furthermore, the damping coefficient associated with the Voigt profile is
\begin{equation}
a_0 = \frac{\Gamma_{\rm L}}{2 \Gamma_{\rm D}} \approx 10^{-3} \left( \frac{T}{10^3 \mbox{ K}} \right)^{-1/2} \left( \frac{m}{2 m_{\rm H}} \right)^{1/2},
\end{equation}
implying that the Voigt and Doppler profiles are essentially identical near line center.  

If pressure broadening is significant, then $\Gamma_{\rm L}$ needs to be replaced by $\Gamma_{\rm L} + \Gamma_{\rm coll}$ in the Lorentz profile, where $\Gamma_{\rm coll}$ is the collisional frequency \citep{mihalas}.  Pressure broadening is similar in nature to Doppler broadening, 
\begin{equation}
\Gamma_{\rm coll} \sim \frac{\lambda_0 \Gamma_{\rm D}}{l_{\rm mfp}} = P \sigma_{\rm H_2} \left( \frac{2}{m k_{\rm B} T} \right)^{1/2},
\end{equation}
where $l_{\rm mfp}$ is the mean free path of collisions and $\sigma_{\rm H_2} \sim 10^{-15}$ cm$^2$ is the cross section of collisions with hydrogen molecules.  Pressure broadening may be ignored if the pressure being sensed is approximately
\begin{equation}
P < 1 \mbox{ mbar} \left( \frac{m}{2 m_{\rm H}} \frac{T}{10^3 \mbox{ K}} \right)^{1/2}.
\end{equation}
We will neglect pressure broadening for our analysis of the sodium lines, while being aware of a shortcoming of our theory: we cannot directly diagnose the value of the \textit{total} pressure being sensed, since it is degenerate with the sodium abundance.  Non-Lorentzian corrections to the sodium line wings are only important far away ($\gtrsim 100 \Gamma_{\rm D}$) from line center and for $P \gtrsim 1$ bar \citep{allard12}.

\begin{figure}
\begin{center}
\vspace{-0.1in}
\includegraphics[width=\columnwidth]{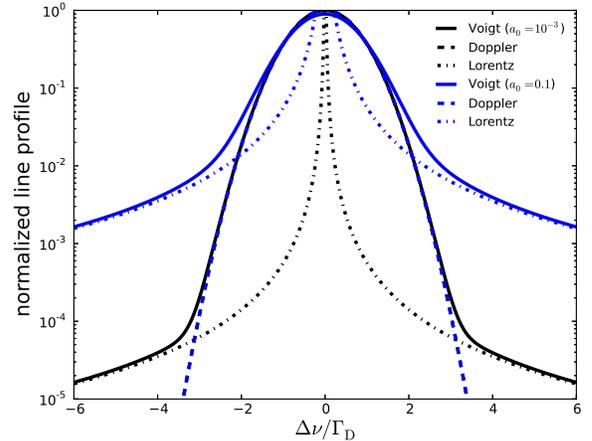}
\end{center}
\vspace{-0.2in}
\caption{Normalized Voigt, Doppler and Lorentz profiles for $a_0 = 10^{-3}$ and 0.1.  The former value is representative of the sodium lines in hot exoplanetary atmospheres, while the latter value is presented as an extreme case.  The line core is well approximated by a Doppler profile out to several Doppler widths.  Beyond this, the line wings are better approximated by a Lorentz profile.  From inspecting these profiles, it is clear that the shape and normalization of the Doppler core are insensitive to temperature, whereas the difference in normalization between it and the Lorentzian wings serves as a temperature diagnostic.}
\label{fig:profiles}
\end{figure}

Consider a point on one of the sodium lines that is separated by $\Delta \lambda$, in wavelength, from line center.  In frequency, the separation is $\Delta \nu$.  If $\Delta \nu = \Gamma_{\rm D}$, then we have
\begin{equation}
\Delta \lambda = \frac{\lambda^2 \Delta \nu}{c} \approx 0.06 \mbox{ \AA} \left( \frac{T}{10^3 \mbox{ K}} \right)^{1/2} \left( \frac{m}{2 m_{\rm H}} \right)^{-1/2},
\end{equation}
implying that the lines are tens of Doppler widths wide.  Even by moving 0.3 \AA~ away from line center, one is already well into the Lorentzian wings of the profile.  In the line wings, we may approximate the line shape with a Lorentz profile.  

Collectively, the sodium line profile in hot exoplanetary atmospheres ($T \sim 1000$ K) is well-approximated by a Doppler core and Lorentzian wings (Figure \ref{fig:profiles}), which allows us to circumvent the more challenging task of inverting a Voigt profile.

\subsection{Review of Previous Analytical Formulae}

In \cite{fortney05}, the assumptions of hydrostatic equilibrium and an isothermal atmosphere led to an expression for the number density,
\begin{equation}
n = n_{\rm ref} \exp{\left( - \frac{z}{H} \right)},
\end{equation}
where $n_{\rm ref}$ is a reference value of the number density, $z$ is the height above some reference radius $R$, $H \equiv k_{\rm B} T / m g$ is the pressure scale height and $g$ is the surface gravity.  Using Pythagoras's theorem and assuming $z \ll R$, \cite{fortney05} obtained (see Figure \ref{fig:schematic} for the geometry)
\begin{equation}
z \approx \frac{x^2}{2R}.
\end{equation}
It follows that the optical depth of the chord associated with the transit radius is
\begin{equation}
\tau = n_{\rm ref} \sigma \int^{+\infty}_{-\infty} \exp{\left( - \frac{x^2}{2HR} \right)} ~dx = n_{\rm ref} \sigma \left( 2 \pi H R \right)^{1/2},
\label{eq:fortney}
\end{equation}
where $\sigma$ is the absorption cross section.  In hindsight, one could have obtained a quick answer by writing $\tau = n_{\rm ref} \sigma X$ and recognising that $X \sim \sqrt{HR}$ is roughly the geometric mean of two vastly different length scales, $H$ and $R$.

Several properties associated with the formula of \cite{fortney05}, restated in our equation (\ref{eq:fortney}), are worth emphasizing.  First, it can only be applied to isothermal situations.  Second, $R$ is a \textit{wavelength-dependent} quantity.  Third, since one is sensing different values of $R$ across wavelength, the value of $n_{\rm ref}$ sampled varies as well.  The number density can be fixed by writing $n_{\rm ref} = n^\prime_{\rm ref} \exp{(-z/H)}$, where $n^\prime_{\rm ref}$ is its value in the line wings.  Unlike $n_{\rm ref}$, $n^\prime_{\rm ref}$ is a wavelength-independent quantity.

The third line of reasoning allowed \cite{lec08} to invert the expression for $\tau$ and obtain
\begin{equation}
z = H \ln{\left[ \frac{n^\prime_{\rm ref} \sigma \left( 2 \pi H R \right)^{1/2}}{\tau} \right]}.
\end{equation}
Since the transit radius always picks out $\tau \sim 1$ by definition and $T$ is not a wavelength-dependent quantity, it follows that
\begin{equation}
T = \frac{mg}{k_{\rm B}} \frac{\partial z}{\partial \lambda} \left[ \frac{\partial \left(\ln\sigma\right)}{\partial \lambda} + \frac{1}{2} \frac{\partial \left(\ln{R}\right)}{\partial \lambda} \right]^{-1}.
\label{eq:t_lec}
\end{equation}
The second term within the square brackets is typically smaller than the first, i.e., $\partial (\ln{R}) / \partial \lambda \ll \partial (\ln{\sigma}) / \partial \lambda$, since we expect the transit radius to vary by fractions of a percent, whereas the cross section is expected to vary by orders of magnitude.  Dropping the second term reproduces equation (2) of \cite{lec08}.  Equation (\ref{eq:t_lec}) is directly applicable to situations where $\partial z / \partial \lambda$ is constant (e.g., Rayleigh scattering), but is less amenable to analyzing spectral line shapes, where $\partial z / \partial \lambda$ is non-constant by definition.

\subsection{Obtaining Temperature and Density in the Isothermal Limit}

\begin{figure}
\begin{center}
\includegraphics[width=\columnwidth]{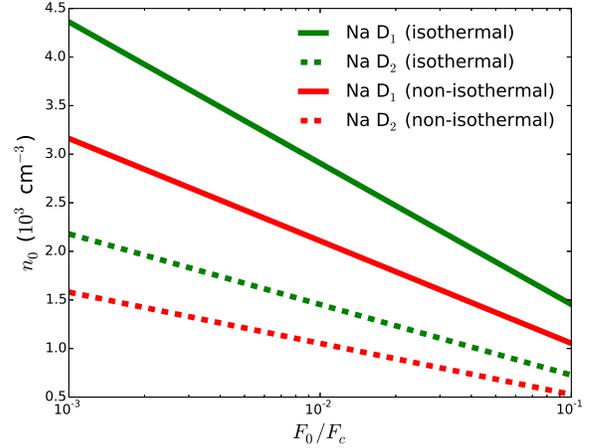}
\includegraphics[width=\columnwidth]{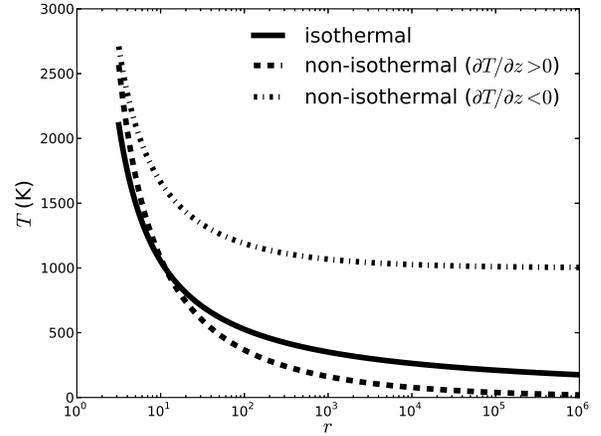}
\end{center}
\vspace{-0.2in}
\caption{Sodium line diagnostics adopting $m=2m_{\rm H}$, $g = 10^3$ cm s$^{-2}$, $R_0 = 10^{10}$ cm, $\Delta R = 10^8$ cm, $\zeta=0.5$ and $T^\prime = 1$ K km$^{-1}$ for illustration.  Top panel: the line-center number density.  Bottom panel: the line-wing temperature.  For the non-isothermal cases, we set $b=2.5$ ($\partial T / \partial z > 0$) and 3.5 ($\partial T / \partial z < 0$).}
\label{fig:diagnostics}
\end{figure}

At line center, the absorption cross section of the sodium atom is \citep{draine11}
\begin{equation}
\sigma = \frac{e^2 f_{\rm lu} \lambda_0}{m_e c} \left( \frac{\pi m}{2 k_{\rm B} T} \right)^{1/2},
\end{equation}
where $f_{\rm lu}$ is the oscillator strength.  The mass and charge of the electron are given by $m_e$ and $e$, respectively.

By denoting the line-center value of $W_\lambda$ as $W_{\lambda_0}$, we may express the line-center optical depth as
\begin{equation}
\tau_0 = - \ln{\left( 1 - W_{\lambda_0} \right)} = -\ln{\left( \frac{F_0}{F_c} \right)},
\end{equation}
where $F_0$ and $F_c$ are the flux at line center and of the continuum, respectively.  For the remainder of the study, we shall deal only with the absorption depth as an observable.  It follows that the number density associated with line center is
\begin{equation}
n_0 = - \ln{\left( \frac{F_0}{F_c} \right)} ~\frac{m_e c}{\pi e^2 f_{\rm lu} \lambda_0} \left( \frac{g}{R_0} \right)^{1/2}.
\label{eq:n0_isothermal}
\end{equation}

In the line wings, the optical depth is given by
\begin{equation}
\tau = n_0 \exp{\left( \frac{\Delta R}{H} \right)} \left( 2 \pi H R \right)^{1/2} \frac{\pi e^2 f_{\rm lu} \Phi}{m_e c},
\label{eq:tau_isothermal}
\end{equation}
with the line shape ($\Phi$) being approximated by a Lorentz profile sufficiently far away from line center,
\begin{equation}
\Phi = \frac{A_{21} \lambda^2 \lambda^2_0}{4 \pi c^2 \left( \lambda - \lambda_0 \right)^2}.
\end{equation}
The transit radius at line center ($R_0$) and in the line wings ($R$) are related by
\begin{equation}
R_0 = R + \Delta R.
\end{equation}
Like $R$, $\Delta R$ is a wavelength-dependent observable.  Equation (\ref{eq:tau_isothermal}) can be inverted to solve for the temperature,
\begin{equation}
T = \frac{m g ~\Delta R}{k_{\rm B} \ln{r}},
\label{eq:temp_isothermal}
\end{equation}
by eliminating $n_0$ using equation (\ref{eq:n0_isothermal}).  We have defined the diagnostic ratio as
\begin{equation}
\begin{split}
r \equiv& \frac{\ln{\left( F/F_c \right)}}{\ln{\left( F_0/F_c \right)}} \left( \frac{\pi m R_0}{2 k_{\rm B} R T} \right)^{1/2} \frac{4 c^2 ~\Delta \lambda^2}{\lambda^3_0 A_{21}} \\
\approx& 2 \times 10^3 ~\frac{\ln{\left( F/F_c \right)}}{\ln{\left( F_0/F_c \right)}} \left( \frac{R_0}{R} \right)^{1/2}\\
&\times \left( \frac{T}{10^3 \mbox{ K}} \right)^{-1/2} \left( \frac{m}{2 m_{\rm H}} \right)^{1/2} \left( \frac{\Delta \lambda}{0.1 \mbox{ \AA}} \right)^2,
\end{split}
\label{eq:diag_ratio}
\end{equation}
where $\Delta \lambda \equiv \lambda - \lambda_0$. 

The temperature is mostly controlled by the difference in transit radii between the line center and wings.  Formally, $\ln{r}$ has a very weak dependence on $T$, so equation (\ref{eq:temp_isothermal}) is not an explicit formula for $T$.  In practice, one sets $T \sim 1000$ K, measures or assumes all of the other parameter values, uses equation (\ref{eq:temp_isothermal}) to compute an updated value of $T$ and iterates.  A converged answer should obtain within a few iterations.

Figure \ref{fig:diagnostics} shows calculations of $n_0$ as a function of $F_0/F_c$ for both sodium lines.  For the sodium D$_1$ line, we have used $\lambda_0 = 5897.558$ \AA~ and $f_{\rm lu} = 0.320$; for the sodium D$_2$ line, we have used $\lambda_0 = 5891.582$ \AA~ and $f_{\rm lu} = 0.641$ \citep{draine11}.  The D$_1$ and D$_2$ lines are sensing different altitudes within the atmosphere.  Figure \ref{fig:diagnostics} also shows the line-wing temperature as a function of the diagnostic ratio.

\subsection{Obtaining Temperature and Density in the Non-isothermal Case}

We generalise to a non-isothermal situation by considering a linear series expansion of the temperature,
\begin{equation}
T = T_{\rm ref} + \frac{\partial T}{\partial z} z,
\end{equation}
where $T_{\rm ref}$ is a reference temperature and we allow the temperature gradient to be constant and either positive or negative,
\begin{equation}
\frac{\partial T}{\partial z} = \pm T^\prime,
\end{equation}
where $T^\prime$ is always a positive number.  We begin with hydrostatic balance,
\begin{equation}
\frac{\partial P}{\partial z} = - \rho g,
\end{equation}
where $P$ is the pressure and $\rho = nm$ is the mass density.  The physical assumption being made is that the buffer or inert gas (e.g., molecular hydrogen) is in hydrostatic equilibrium and the collisional time between itself and the sodium atoms is so short that the latter are effectively coupled to the former as a single fluid.

By using the ideal gas law, $P = n k_{\rm B} T$, one obtains
\begin{equation}
\int^n_{n_{\rm ref}} \frac{1}{n} ~dn = - \frac{m \tilde{g}}{k_{\rm B}} \int^z_0 \frac{1}{T_{\rm ref} \pm T^\prime z} ~dz.
\end{equation}
Here, $\tilde{g} \equiv g \pm k_{\rm B} T^\prime/m$ functions like an ``effective gravity" with an additional term associated with the constant temperature gradient.  Completing the integration yields
\begin{equation}
n = 
\begin{cases}
n_{\rm ref} \left( 1 + \frac{T^\prime z}{T_{\rm ref}} \right)^{-\left(b+1\right)}, & ~\frac{\partial T}{\partial z} > 0, \\
n_{\rm ref} \left( 1 - \frac{T^\prime z}{T_{\rm ref}} \right)^{b-1}, & ~\frac{\partial T}{\partial z} < 0, \\
\end{cases}
\label{eq:numdens}
\end{equation}
where the non-isothermal pressure scale height is $H \equiv T_{\rm ref}/T^\prime$ and the index,
\begin{equation}
b \equiv \frac{mg}{k_{\rm B} T^\prime},
\end{equation}
is the ratio of the non-isothermal to the isothermal scale heights.

For $\partial T / \partial z < 0$, the minimum value of $b$ is set by the adiabatic lapse rate such that $b \ge (2+N_{\rm dof})/2$, where $N_{\rm dof}$ is the number of degrees of freedom of the buffer gas.  For a diatomic gas, we have $N_{\rm dof}=5$ and $b \ge 3.5$ (ignoring the vibrational modes).

The chord optical depth is
\begin{equation}
\tau = n_{\rm ref} \sigma \left( \frac{2 R T_{\rm ref}}{T^\prime} \right)^{1/2} \int^{+\pi/2}_{-\pi/2} \left(\cos\theta\right)^{N_b} ~d\theta,
\label{eq:tau_chord_nonisothermal}
\end{equation}
where the index $N_b$ is
\begin{equation}
N_b = 
\begin{cases}
2b, & ~\frac{\partial T}{\partial z} > 0, \\
2b-1, & ~\frac{\partial T}{\partial z} < 0. \\
\end{cases}
\end{equation}
The integral in equation (\ref{eq:tau_chord_nonisothermal}) has no general analytical solution.  However, it is soluble if we consider integer values of $N_b$.  This is not unreasonable.  For example, for HD 189733b \cite{huitson12} estimate $T^\prime \approx 1$ K km$^{-1}$, which yields $N_b \approx 5$ for $\partial T / \partial z > 0$.  For $N_b=0, 2, 4$ and 6, the integral is $\pi$, $\pi/2$, $3\pi/8$ and $5\pi/16$, respectively.  Essentially, the integral may be represented as $\zeta \pi$, where $\zeta \approx 0.1$--1.  Numerical evaluations of the integral, over a continuous range of values of $N_b$, confirm our intuition (not shown).  The optical depth thus becomes
\begin{equation}
\tau = \zeta \pi n_{\rm ref} \sigma \left( \frac{2 R T_{\rm ref}}{T^\prime} \right)^{1/2},
\end{equation}
similar in spirit to its isothermal counterpart where we again have the characteristic length scale being $X \sim \sqrt{HR}$.  Generalising from the isothermal situation, we note that $n_{\rm ref}$, $R$ and $T_{\rm ref}$ are wavelength-dependent quantities.

At line center, we again assume a Doppler profile and derive the number density,
\begin{equation}
n_0 = - \ln{\left( \frac{F_0}{F_c} \right)} ~\frac{m_e c}{\zeta \pi e^2 f_{\rm lu} \lambda_0} \left( \frac{k_{\rm B} T^\prime}{\pi m R_0} \right)^{1/2},
\label{eq:n0}
\end{equation}
an expression which holds for both $\partial T / \partial z > 0$ and $\partial T / \partial z < 0$.  Unlike before, $n_0$ now depends also on $T^\prime$.  In the line wings, we can again assume a Lorentz profile, eliminate $n_0$ and obtain
\begin{equation}
T = 
\begin{cases}
T^\prime \Delta R \left[ r^{1/\left( b+1 \right)} - 1 \right]^{-1}, & ~\frac{\partial T}{\partial z} > 0, \\
T^\prime \Delta R \left[ 1 - r^{-1/\left( b-1 \right)} \right]^{-1}, & ~\frac{\partial T}{\partial z} < 0. \\
\end{cases}
\label{eq:T}
\end{equation}
The definition for $r$ is identical to the one stated in equation (\ref{eq:diag_ratio}).  It is worth noting that our expression for $T$, in equation (\ref{eq:T}), does not depend on $\zeta$.

Equations (\ref{eq:n0}) and (\ref{eq:T}) constitute an under-determined system: two equations and three unknowns ($n_0$, $T$, $T^\prime$).  We will break this degeneracy by using measurements from both sodium lines to determine $T^\prime$.

\section{Application to Data}

To apply the diagnostics to data requires that both sodium lines are at least partially resolved, such that for each line we may measure the absorption depths at line center and in the line wing.  Generally, at the same distance ($\Delta \lambda$) from line center, the line wings of the D$_1$ and D$_2$ sodium lines are sampling different temperatures, which we denote by $T_1$ and $T_2$, respectively.  We denote the line-center temperatures by $T_{0,1}$ and $T_{0,2}$.
\begin{enumerate}

\item To determine if the system is isothermal, one computes the ratio of the line-wing temperatures,
\begin{equation}
\frac{T_1}{T_2} = \frac{\Delta R_1}{\Delta R_2},
\end{equation}
where $\Delta R_1$ and $\Delta R_2$ are the differences in transit radii, between the line center and wing, for the D$_1$ and D$_2$ lines, respectively.  To a very good approximation, the preceding expression holds regardless of whether one uses the isothermal or non-isothermal formula for $T$.  If $T_1 = T_2$, the isothermal formulae in equations (\ref{eq:n0_isothermal}) and (\ref{eq:temp_isothermal}) will suffice.

\item If $T_1 \ne T_2$, then we first use the isothermal formula in equation (\ref{eq:temp_isothermal}) to estimate $T_1$, $T_2$ and 
\begin{equation}
\frac{\partial T}{\partial z} = \frac{T_1 - T_2}{R_1 - R_2}.
\end{equation}
After assessing if the temperature gradient is positive or negative, we use the appropriate non-isothermal formula in equation (\ref{eq:T}) to iterate for $T_1$, $T_2$ and $T^\prime$.

\item With the value of $T^\prime$ in hand, we may also estimate the number densities using equations (\ref{eq:numdens}) and (\ref{eq:n0}).

\end{enumerate}

We apply our diagnostics to the sodium doublet measured in HD 189733b using HARPS ground-based data \citep{wy15}.  At line center, the absorption depths and transit radii are
\begin{equation}
\begin{split}
F_{0,1}/F_c &= 0.9960 \pm 0.0007, ~R_{0,1}/R_{\rm J} = 1.227 \pm 0.021,\\
F_{0,2}/F_c &= 0.9936 \pm 0.0007, ~R_{0,2}/R_{\rm J} = 1.277 \pm 0.020.
\end{split}
\end{equation}
At $\Delta \lambda = 0.3$ \AA~ from line center, the absorption depths and transit radii are
\begin{equation}
\begin{split}
F_1/F_c &= 0.9987 \pm 0.0007, ~R_1/R_{\rm J} = 1.168 \pm 0.022,\\
F_2/F_c &= 0.9974 \pm 0.0007, ~R_2/R_{\rm J} = 1.196 \pm 0.022.
\end{split}
\end{equation}
All quantities were extracted using Gaussian fits to the data.  The white-light radius used to normalise the transit radii is $(1.138 \pm 0.027) R_{\rm J}$ \citep{torres08}.  We verified that instrumental broadening is not a major source of error \citep{wy15}.

To faciliate comparison with \cite{huitson12}, we use $g = 2141$ cm s$^{-2}$ and $m = 2.3 m_{\rm H}$ for HD 189733b.  We propagate the data uncertainities in quadrature.  Using the isothermal formula, we obtain $T_1 = 3117 \pm 79$ K and $T_2 = 4240 \pm 103$ K.  Next, we derive $T_1/T_2 = 0.7284 \pm 0.0256$, which indicates that the non-isothermal treatment is warranted.  We obtain
\begin{equation}
\frac{\partial T}{\partial z} = 0.4376 \pm 0.0154 \mbox{ K km}^{-1},
\end{equation}
thus confirming the positive temperature gradient reported by \cite{huitson12}.  Our line-wing and line-center temperatures are 
\begin{equation}
\begin{split}
T_1 &= 2460 \pm 86 \mbox{ K}, ~T_{0,1} = 4306 \pm 151 \mbox{ K}, \\
T_2 &= 3336 \pm 117 \mbox{ K}, ~T_{0,2} = 5870 \pm 206 \mbox{ K}.
\end{split}
\end{equation}
The number densities are 
\begin{equation}
\begin{split}
n_1 &= \left( 1.439 \pm 0.051 \right) \times 10^4 \mbox{ cm}^{-3}, \\
n_2 &= \left( 1.219 \pm 0.043 \right) \times 10^4 \mbox{ cm}^{-3},\\
n_{0,1} &= 3.990 \pm 0.140 \mbox{ cm}^{-3}, \\
n_{0,2} &= 3.128 \pm 0.110 \mbox{ cm}^{-3},
\end{split}
\end{equation}
while $b = 13.63 \pm 0.48$.  If the abundance of sodium is solar, then the total pressure implied is $\lesssim 5$ nbar.

The temperatures and temperature gradient inferred using our analytical diagnostics are in broad agreement with the numerical models of \cite{wy15}, applied to the same data set.

\section{Conclusion}

We have presented a previously-unelucidated theory for inferring the temperatures, temperature gradient and number densities associated with the absorption depths and transit radii of the sodium doublet lines.  Our formalism is applicable to other alkali metal lines such as potassium.  Applying our diagnostics to infrared lines of molecules requires the inclusion of pressure broadening.

\acknowledgments
K.H., A.W., B.L., D.E. and C.L. acknowledge support from the PlanetS National Center of Competence in Research (NCCR) of the Swiss National Science Foundation.  D.K.S. acknowledges funding from the European Research Council grant agreement n$^{\circ}$ 336792 (FP7/2007--2013).


\label{lastpage}

\end{document}